\documentclass[twocolumn]{aa}  
\usepackage{graphicx}
\usepackage{txfonts}
\usepackage{xcolor}
\definecolor{xlinkcolor}{cmyk}{1,0.6,0,0}
\usepackage[breaklinks=true,
   bookmarks=false,         % show bookmarks bar?
    pdfnewwindow=true,      % links in new window
   colorlinks=true,    % false: boxed links; true: colored links
  linkcolor=xlinkcolor,     % color of internal links
     citecolor=xlinkcolor,     % color of links to bibliography
     filecolor=xlinkcolor,  % color of file links
     urlcolor=xlinkcolor,      % color of external links
final=true
]{hyperref}

\begin{document}

   \title{The isolated dark matter-poor galaxy that ran away}

   \subtitle{An example from IllustrisTNG}

   \author{Ana Mitra{\v s}inovi{\' c}\inst{1}
          \and
          Majda Smole\inst{1}
          \and 
          Miroslav Micic\inst{1}
          }

   \institute{$^1$Astronomical Observatory, Volgina 7, 11060 Belgrade, Serbia\\
              \email{amitrasinovic@aob.rs}
             }

   \date{Received xxx; accepted xxx}

% \abstract{}{}{}{}{} 
% 5 {} token are mandatory
 
  \abstract
  {Since the discovery of dark matter-deficient galaxies, numerous studies have shown that these exotic galaxies naturally occur in the $\Lambda$CDM model due to stronger tidal interactions. They are typically satellites, with stellar masses in the $10^8-10^9\;\mathrm{M}_\odot$ range, of more massive galaxies. The recent discovery of a massive galaxy lacking dark matter and also lacking a more massive neighbor is puzzling. Two possible scenarios have been suggested in the literature: either the galaxy lost its dark matter early or it had been lacking ab initio. As a proof of concept for the former assumption, we present an example from IllustrisTNG300. At present, the galaxy has  a stellar mass of $M_\star \simeq 6.8 \cdot 10^9\; \mathrm{M}_\odot$,  with no gas, $M_\mathrm{DM}/M_\mathrm{B} \simeq 1.31$, and a stellar half-mass radius of $R_{0.5,\star} = 2.45\;\mathrm{kpc}$. It lost the majority of its dark matter early, between $z = 2.32$ and $z = 1.53$. Since then, it has continued to dwell in the cluster environment, interacting with the cluster members without merging, while accelerating on its orbit. Eventually, it left the cluster and it has spent the last $\sim 2\;\mathrm{Gyr}$ in isolation, residing just outside the most massive cluster in the simulation. Thus, the galaxy represents the first example found in simulations of both an isolated dark matter-poor galaxy that lost its extended envelope early and a fairly compact stellar system that has managed to escape.}


   \maketitle
%
%-------------------------------------------------------------------

\section{Introduction}

The discovery of galaxies lacking dark matter \citep{vanDokkum2018,vanDokkum2019} has sparked a debate on recent challenges to the standard cosmological model ($\Lambda$CDM). \citet{Yang2020PhRvL} argued that self-interacting dark matter is a favorable scenario for explaining the formation and existence of such galaxies. Questioning the uncertainties of the data, modified Newtonian dynamics (MOND) has also been proposed \citep[e.g.,][]{Famaey+2018MNRAS.480..473F,Haghi+2019MNRAS.487.2441H}. Also, \citet{Martin+2018ApJ...859L...5M}  questioned the validity of the data, arguing that the discovered galaxy is a typical dwarf, but this claim was rejected \citep[e.g.,][]{vanDokkum+2018RNAAS...2...54V,Wasserman+2018ApJ...863L..15W,Keim+2022ApJ...935..160K}. Furthermore, \citet{Saulder+2020MNRAS.491.1278S}  warned that isolated dark matter-deficient galaxies in simulations could represent a numerical artifact (arising from problems with the algorithm identifying bound structures and the periodic boundary conditions of the simulation boxes).

Numerous studies have since shown that these exotic galaxies are not in tension with the $\Lambda$CDM model and that they may naturally occur due to stronger tidal interactions \citep[e.g.,][]{ogiya2018,montes2020,shin2020,jackson2021,maccio2021,Ogiya2022,trujillogomez2021}. In particular, \citet{moreno2022} estimated that $\sim 30\%$ massive central galaxies have at least one dark matter-deficient satellite with stellar masses in the $10^8-10^9\;\mathrm{M}_\odot$ range. Exploring the evolution of satellite galaxies in IllustrisTNG simulation boxes, \citet{Montero-Dorta+2022arXiv221212090M} also considered dark matter-deficient systems and reinforced the notion that the tidal stripping is responsible for the lack of dark matter.

In this context, the recent discovery \citep{Cameron+2023A&A} of a massive NGC 1277 galaxy lacking dark matter, while also lacking a more massive companion, was puzzling. These authors considered multiple possible scenarios to explain this phenomenon:\ either the galaxy lost its dark matter early or it had been lacking ab initio. To test the viability of the former assumption,  we set our aims on the search for candidates in IllustrisTNG. From the theoretical point of view, the existence of backsplash galaxies \citep[][and references within]{Borrow+2023MNRAS.520..649B}, namely, ones that had resided inside a cluster and have since migrated back outside, would favor the possibility that galaxies  may, in fact, "run away" from dense environments. Moreover, \citet{Chilingarian+Zolotukhin2015Sci...348..418C} argued that the observed isolated compact elliptical and isolated quiescent dwarf galaxies are tidally stripped systems that had run away from their host clusters.

Here, we present one such example from IllustrisTNG300: an isolated dark matter-poor galaxy that lost its extended dark matter envelope early. Since then, it has dwelled in the cluster environment, interacting with the cluster members without merging, while accelerating on its orbit; eventually, it left the cluster, spending the last $\sim2$ Gyr in isolation. This paper is organized as follows. We describe our candidate selection criteria in Section~\ref{sec:candidate}. In Section~\ref{sec:results}, we present the evolutionary path of the galaxy separated into stages. We discuss our results, along with the limitations and caveats of the simulation in Section~\ref{sec:discussion}. We summarize our work and give our concluding remarks in Section~\ref{sec:summary}.
   
\section{Candidate selection} \label{sec:candidate}

We utilized the IllustrisTNG cosmological simulation suite\footnote{Publicly available at \url{https://www.tng-project.org/data/}.} \citep{TNGmethods2017,TNGmethods2018,Nelson+2019ComAC}. Choosing an appropriate simulation box was not a straightforward task. The largest simulation box, TNG300 \citep{Marinacci+2018,Naiman+2018,Nelson+2018,Pillepich+2018,Springel+2018}, has the poorest particle mass resolution. However, due to its size, it allows us to explore galaxies experiencing vastly different environments. Knowing that dark, matter-poor galaxies are not always low-mass objects  \citep[e.g.,][]{Cameron+2023A&A}, the resolution limit should not pose a challenge. Moreover, the TNG300 box offers the largest sample of galaxies, which would help identify extremely rare present-day isolated dark, matter-poor galaxies.

We describe our selection criteria in the following. First, we considered only the subhalos of cosmological origin (that have formed due to the process of structure formation and collapse, and have $\texttt{SubhaloFlag}=1$\footnote{Meaning that a subhalo should be considered as a galaxy. Otherwise, subhalos with $\texttt{SubhaloFlag}=0$ are, at their formation time, satellites, tidal debris, or kinematically distinct substructures of already formed galaxies. As such, they typically have little to no dark matter and would contaminate our search.}). Then, at $z=0$, we extracted the ones that are isolated (i.e., they are the only subhalo within their parent halo) and that have the dark-matter-to-baryon-mass ratio of $M_\mathrm{DM}/M_\mathrm{B}<2$. While the filter yielded 448 subhalos, the majority of them are poorly resolved. We chose the best-resolved, most massive subhalo, ID3254912\footnote{\texttt{SubfindID} at $z=0$.} (herein, "the galaxy"). The galaxy has, at $z=0$, a total mass of $M \simeq 1.56 \cdot 10^{10}\; \mathrm{M}_\odot$, with the mass of stellar component of $M_\star \simeq 6.8 \cdot 10^9\; \mathrm{M}_\odot$, no gas, and $M_\mathrm{DM}/M_\mathrm{B} \simeq 1.31$. The respective half-mass radii are $R_{0.5,\mathrm{DM}} = 2.95\;\mathrm{kpc}$ and $R_{0.5,\star} = 2.45\;\mathrm{kpc}$\footnote{Throughout this work, all given units are physical, for convenience.}.

\section{Results} \label{sec:results}

While the galaxy still contains a non-negligible amount of dark matter at present, the total mass of the dark matter component is comparable to its stellar counterpart, indicating it is a dark matter-poor galaxy. Additionally, about half of the total dark matter mass is confined within the stellar half-mass radius, suggesting a history of substantial stripping. Due to the outside-in nature of the tidal stripping \citep[e.g.,][]{Diemand+2007,Choi+2009}, the particles residing in the core could not be stripped without a significant disruption of the stellar component and its mass-loss \citep[e.g.,][]{Smith+2016,Lokas2020}. In the following, we present and discuss the galaxy's cosmological evolution according to  its subsequent stages.

\subsection{Mergers and early history} \label{sec:mergers}

The galaxy formed at $z=10$, with $M \simeq 2.6 \cdot 10^9\; \mathrm{M}_\odot$ and $M_\mathrm{DM}/M_\mathrm{B} \simeq 5.88$. Until $z = 7.01$, just before its merger phase, it acquired a substantial amount of dark matter through accretion, resulting in $M \simeq 9.7 \cdot 10^9\; \mathrm{M}_\odot$ and $M_\mathrm{DM}/M_\mathrm{B} \simeq 10.81$.

According to the \texttt{SubLink} \citep{Rodriguez-Gomez+2015MNRAS} data, this galaxy merged with 16 subhalos. The merger phase spanned over 17 early snapshots and lasted for $1.84\;\mathrm{Gyr}$. The galaxy had only one major merger, with a mass ratio of $q\simeq 0.599$ at $z=3.71$. However, given that subsequent mergers have mass ratios significantly below $q = 0.1$, these may have been misclassified. They are likely to be accreted non-cosmological subhalos (i.e., clumps). Checking the merger history catalog \citep{Rodriguez-Gomez+2017,Eisert+2023MNRAS} confirms this assumption. The galaxy has had two mergers in total:\ one major and one minor, which were concluded in nearly subsequent snapshots. Considering this and typical merger timescales, these are not likely to be two separate events (i.e., binary mergers) but, rather, a single multiple merger.

After this phase, at $z = 2.44$ (just before the first infall), the galaxy reached $M \simeq 1.17 \cdot 10^{11}\; \mathrm{M}_\odot$, with $M_\mathrm{DM}/M_\mathrm{B} \simeq 6.9$ and with the total masses of sub-components (dark matter, gas, and stars) of: $M_\mathrm{DM} \simeq 1.03 \cdot 10^{11}\; \mathrm{M}_\odot$, $M_\mathrm{G} \simeq 1.43 \cdot 10^{10}\; \mathrm{M}_\odot$, and $M_\star \simeq 5.94 \cdot 10^{8}\; \mathrm{M}_\odot$. The respective half-mass radii are: $R_{0.5,\mathrm{DM}} = 19.71\;\mathrm{kpc}$, $R_{0.5,\mathrm{G}} = 20.82\;\mathrm{kpc}$ and $R_{0.5,\star} = 3.36\;\mathrm{kpc}$. The galaxy has also acquired a black hole particle, with $M_\bullet \simeq 1.82 \cdot 10^{7}\; \mathrm{M}_\odot$.

\subsection{First infall and intense phase} \label{sec:firstinfall}

The galaxy starts its first infall into a small group with $M \simeq 1.92 \cdot 10^{12}\; \mathrm{M}_\odot$, containing a little over 20 subhalos (herein, "members," when referring to subhalos within groups and clusters) at $z = 2.32$. As the infall starts, the galaxy immediately interacts with a group of already interacting galaxies, one of which is the most massive galaxy of the group. In this process, the galaxy abruptly gets stripped of about half its total mass from the outskirts, both dark matter and gas.

\begin{figure}[ht!]
\centering \includegraphics[width=.99\columnwidth, keepaspectratio]{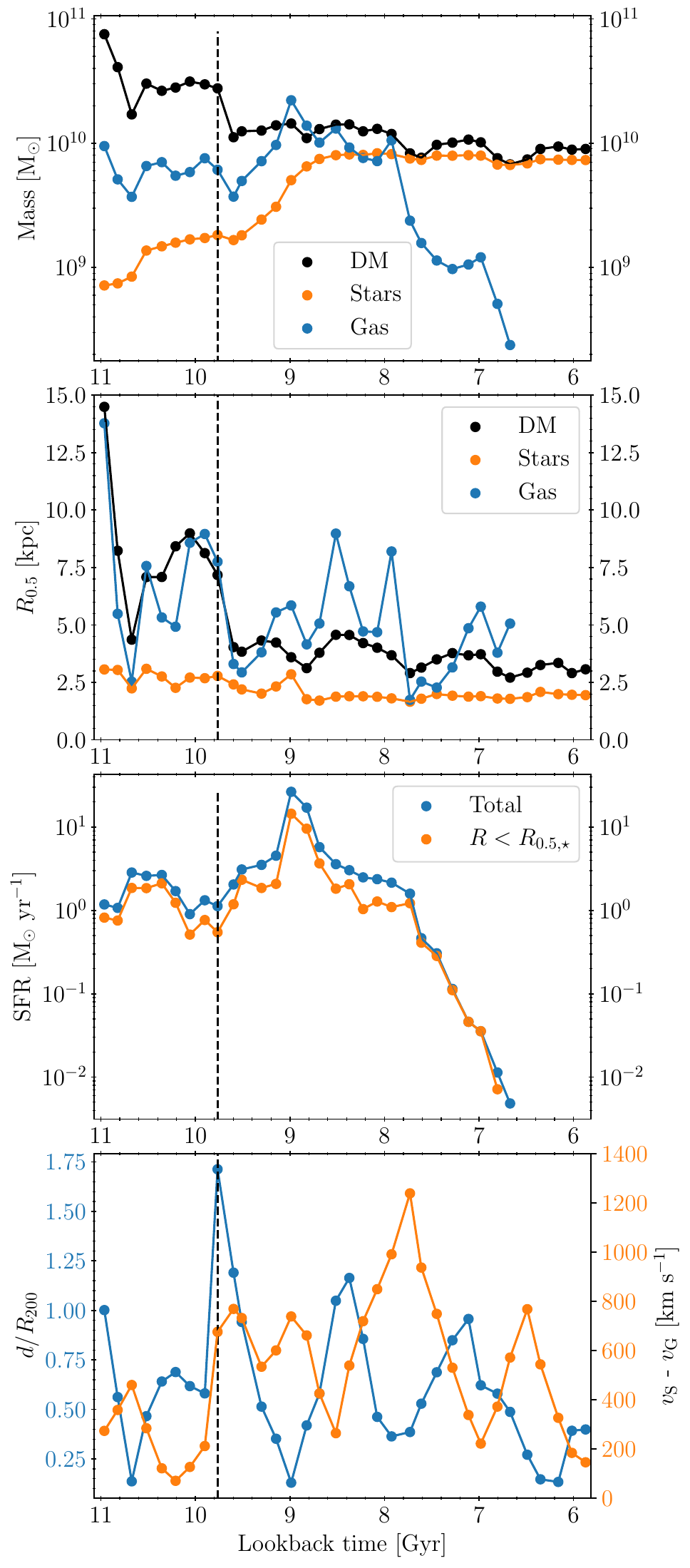}
\caption{Total mass of galactic components (dark matter, stars, gas) shown from top to bottom, with their half-mass radii, $R_{0.5}$ (different colors represent different components, as per the legend); star formation rate (SFR); relative distance from the center of the group or cluster (blue, left $y$-axis), and the galaxy's relative velocity (orange, right $y$-axis). The dashed vertical line represents the moment when the group merges with a nearby cluster.}
\label{fig:intense}
\end{figure}

The evolution of the relevant parameters are shown in Figure~\ref{fig:intense}. Since the galaxy is almost constantly tidally disturbed during this intense phase, as a measure of its center, we used the position of a particle with the lowest potential energy, instead of a calculated center of mass. This provides a more accurate estimate, as the difference between the two can become significant due to tidal structures \citep[e.g.,][]{Mitrasinovic2022SerAJ.204...39M}. Aside from being tidally disturbed, during this period, the galaxy occasionally and briefly becomes a satellite of more massive cluster members as it interacts with them. This can be roughly estimated from the irregularities in its orbit and relative velocity curve.

During its first pericentric passage, the galaxy accretes some dark matter and gaseous particles, and the star formation rate (SFR) increases, producing stars whose total mass consequentially increases. At $z = 1.6$ (snapshot 38 and lookback time of $\simeq 9.77\;\mathrm{Gyr}$), the group merges with a nearby cluster of $M \simeq 1.61 \cdot 10^{13}\; \mathrm{M}_\odot$ containing a little over $100$ members. Upon this event, the galaxy loses the remaining majority of its dark matter (at $z=1.53$), while accreting a considerable amount of gas on its way to the first pericenter ($z=1.3$, snapshot 43, and lookback time of $\simeq 8.99\;\mathrm{Gyr}$). This triggers a burst of star formation, sharply increasing the stellar mass to essentially present-day value. During this period, the dark matter to baryon mass ratio $M_\mathrm{DM}/M_\mathrm{B}$ drops as low as $0.52$. While a fraction of accreted gas is transformed into stars, its majority gets stripped between the second and third pericentric passage, and the galaxy becomes gas-free. As a consequence, $M_\mathrm{DM}/M_\mathrm{B}$ fluctuates around $1$ until the present (i.e., $z=0$).

It is worth mentioning that the galaxy loses its acquired black hole particle near the first pericenter in the cluster. However, this is not a physical process but stems from the black hole re-centering scheme employed in the simulation. The problem was highlighted by \citet{Borrow+2023MNRAS.520..649B} and it is known to be quite common among backsplash galaxies.

\subsection{Cluster dwelling and the isolated phase} \label{sec:cluster}

\begin{figure}[ht!]
\centering \includegraphics[width=\columnwidth, keepaspectratio]{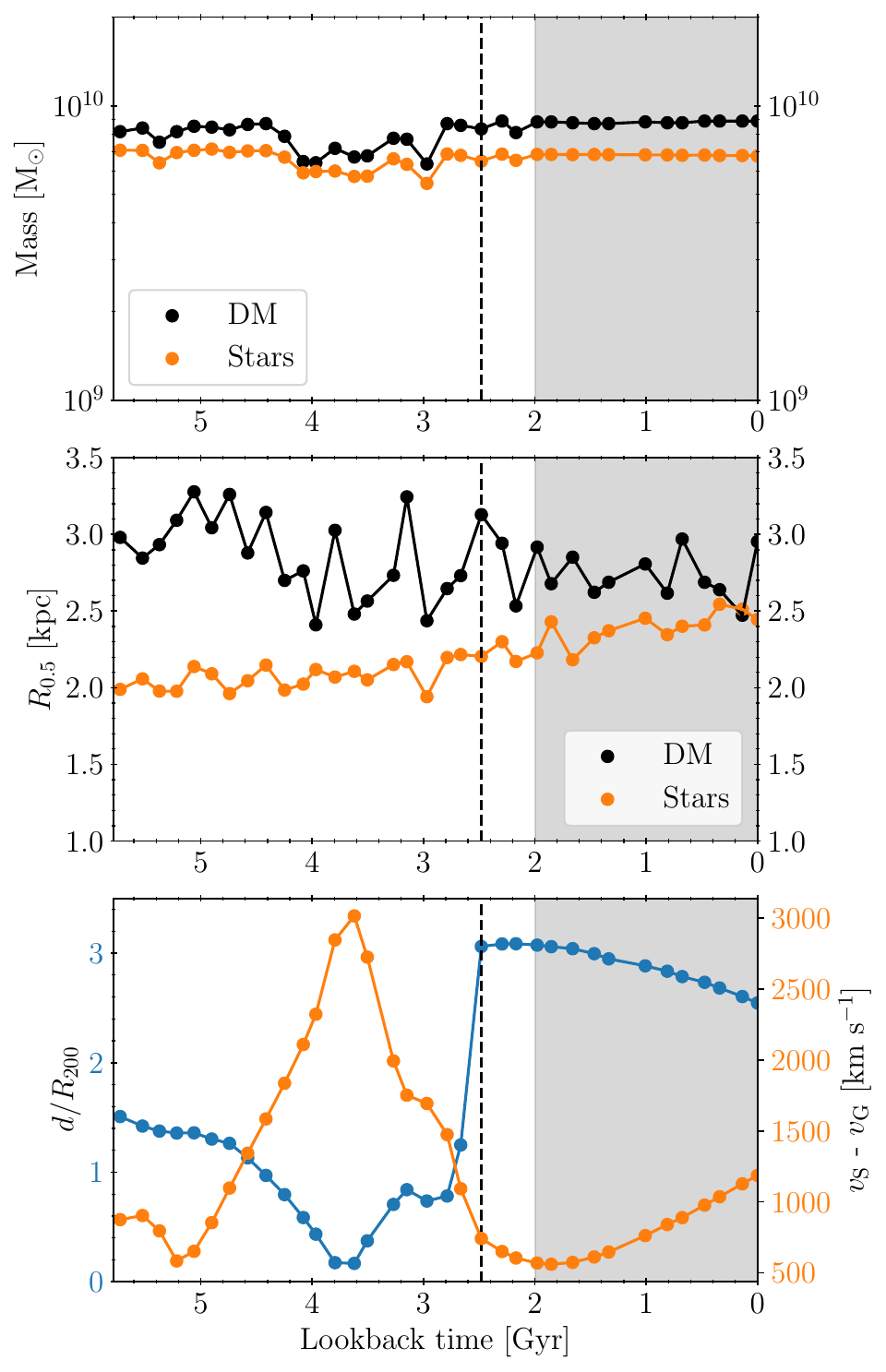}
\caption{Same format and notation as in Figure~\ref{fig:intense}, with the exclusion of gas (and by extent SFR) and the inclusion of a shaded area, representing the time galaxy spends in isolation.}
\label{fig:later}
\end{figure}

We go on to consider the next phase when the host cluster merges with a nearby massive cluster at $z = 0.58$ (snapshot 64 and lookback time of $\simeq 5.72\;\mathrm{Gyr}$). The resulting cluster has $M \simeq 8.25 \cdot 10^{14}\; \mathrm{M}_\odot$ and over $4000$ members. In Figure~\ref{fig:later}, we show the relevant evolution parameters in this later phase. During its time spent in this cluster, the galaxy blazes through, reaching the first and only pericenter around $z=0.31$ (snapshot 77 and lookback time $\simeq 3.62\;\mathrm{Gyr}$) and its relative velocity peaking at $v_\mathrm{S} - v_\mathrm{G} \simeq 3016.95\;\mathrm{km}\;\mathrm{s^{-1}}$. Around the pericenter, the galaxy loses a small fraction of its stellar mass and almost all its dark matter particles at the outskirts.

\begin{figure}[ht!]
\centering \includegraphics[width=.98\columnwidth, keepaspectratio]{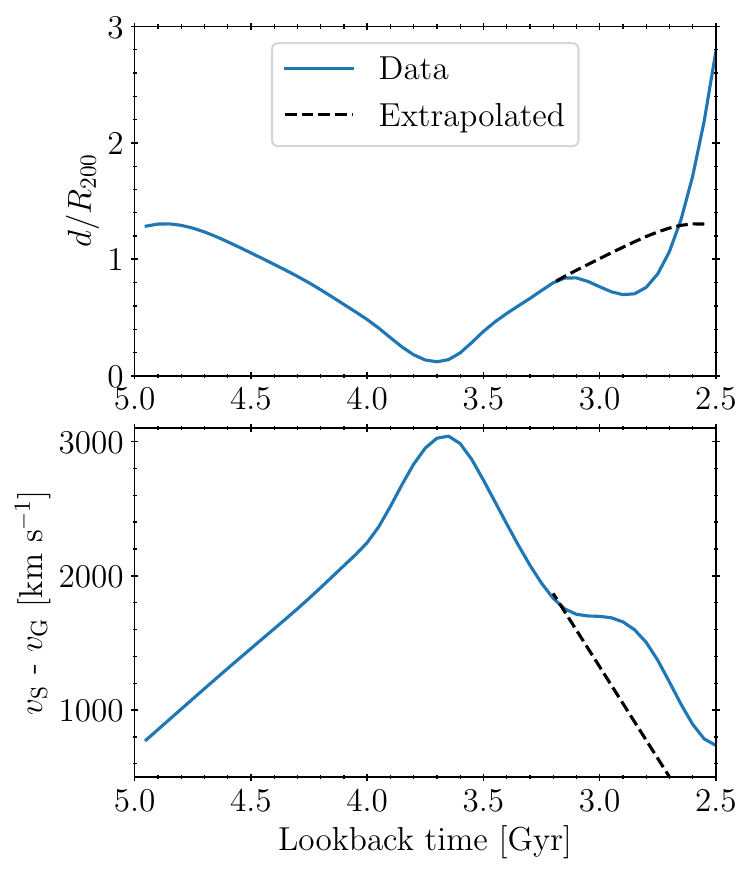}
\caption{Relative distance from the cluster's center (top) and the galaxy's relative velocity (bottom). Solid blue line (data) represents the real values, as seen in Figure~\ref{fig:later}. Dashed black line (extrapolated) represents the data obtained by calculating galaxy's orbit in an idealized smooth cluster potential.}
\label{fig:extrpl}
\end{figure}

On its way to the apocenter, it accretes back some stellar and dark matter particles until it engages in an interaction with multiple cluster members near the cluster outskirts at $d /R_{200}\simeq0.84$. Upon that event, the galaxy suffers mild to moderate tidal stripping, accelerates, diverts its orbit, and gets on its way out of the cluster. This interaction lasts for no more than four consecutive snapshots, making it almost impossible to examine its characteristics in detail. To isolate the interaction's contribution from global cluster effects, we approximate the cluster contribution with smooth NFW potential \citep{nfw1997}, which corresponds to the calculated cluster density profile. We calculated the galaxy's orbit in this idealized smooth cluster potential and the results are shown in Figure~\ref{fig:extrpl}. The orbit is diverted during the interaction, and after it ends, the galaxy gets kicked outside the cluster, as the difference between the realistic and expected distances becomes considerably high. At the same time, the galaxy accelerates, and its real relative velocity compared to the expected values is about twice as high at all times, peaking at $1371\;\mathrm{km}\;\mathrm{s^{-1}}$ when the expected relative velocity is $640\;\mathrm{km}\;\mathrm{s^{-1}}$. This process remarkably resembles the gravitational slingshot mechanism.

While it is still gravitationally bound to the host, the cluster merges with the most massive other cluster in the simulation ($M \simeq 2.64 \cdot 10^{15}\; \mathrm{M}_\odot$ and a little over $12000$ members) at $z=0.2$ (snapshot 84 and lookback time $=2.48\;\mathrm{Gyr}$). The galaxy is located on the opposite side (from where the cluster merger occurs), at a distance of $d/R_{200}\simeq 3.06$ from the new host. Since the cluster merger is not an instantaneous process, in a few subsequent snapshots, we can see that the galaxy is still gravitationally bound to its previous host, until it gets detected as isolated at $z=0.15$ (snapshot 87 and lookback time $\simeq 2\;\mathrm{Gyr}$).

Since for $\sim 2\;\mathrm{Gyr}$, it is dwelling in isolation, just outside the most massive cluster in the simulation box, with essentially no changes in its mass and with its stellar half-mass radius slightly stretching. Chaotic variations in the dark matter half-mass radius, $R_{0.5,\mathrm{DM}}$, are due to particle resolution. The number of dark matter particles fluctuates around $140$, as the galaxy constantly captures and loses single particles at its outskirts. The present-day stellar photometric data for SDSS bands, in particular $g$ and $r$, are taken from \citet{Nelson+2018}. With $(g-r) = (-23.153 + 24.002) \simeq 0.85$, the galaxy is quite red and almost perfectly spherical, according to the stellar mass tensor obtained from the supplementary catalog \citep{Genel+2015ApJ...804L..40G}. An independent calculation of the stellar mass tensor, going back to the time when the stellar component becomes well-resolved for such task, confirms that the galaxy remains almost perfectly spherical at all times. The result is expected, as the stars form at the very core of a tidally stripped dark matter halo.

\section{Discussion} \label{sec:discussion}

The galaxy presented in this work has had quite an interesting evolution, which cannot be considered typical in any way. With $M_\mathrm{DM}/M_\mathrm{B} \simeq 1.31$, it certainly is dark matter-poor, but the dark matter still accounts for a little over $50\%$ of the total mass (with a large fraction of it confined within the stellar half-mass radius). This discrepancy with the observed dark matter-deficient galaxies could be because simulated galaxies typically have higher central dark matter fractions \citep[e.g.,][]{Lovell+2018MNRAS.481.1950L}, which are not easily stripped. Particle resolution can also play a role and this mass ratio should thus be considered an upper limit.

While its present-day color, shape, and extent make it easy to argue that the galaxy should be classified as a spherical dwarf (dSph), its stellar component is overly massive for a dwarf galaxy \citep{Revaz+Jablonka2018A&A...616A..96R}. Hence, we may claim the galaxy is a massive compact galaxy (MCG), but this would not satisfy the compact observational criterion given by \citet{Barro+2013ApJ...765..104B} or a more liberal one given by \citet{Damjanov+2015ApJ...806..158D}, nor is it massive enough overall \citep[e.g.,][]{Lohmann+2023MNRAS.524.5266L}. Given that almost the entirety of its stellar component is $\sim 9\;\mathrm{Gyr}$ old and it has largely remained unchanged since its formation, the galaxy can represent an example of a so-called relic galaxy \citep{Ferre-Mateu+2015ApJ...808...79F}. Observationally confirmed relics (also known as red nuggets) are typically more massive than our example galaxy \citep{Lisiecki+2023A&A...669A..95L}, but this assumption may be due to bias.

Interestingly, since the galaxy lost the majority of its dark matter very early, it does not represent a tidally stripped remnant of a massive galaxy. Instead, its present-day stellar component formed in a single, brief starburst episode at the core of an already stripped dark matter halo. Even though the galaxy does not satisfy the criteria for compact-size galaxies, we argue that the early loss of an extended dark matter and the gaseous envelope has made it compact enough to avoid mergers in higher-density environments. Moreover, since the stripping happened very early on, massive galaxy clusters had not yet formed. This, in addition to its fairly compact size, has helped the galaxy avoid mergers and limited its interactions to members of proto-clusters and later clusters through non-merger events, as it has continued to accelerate on its orbit. After one such interaction that appears to have played out as essentially a gravitational slingshot, the galaxy left the most massive cluster in the simulation. This is the first example found in simulations of both an isolated dark matter-poor galaxy that lost its extended envelope early and a fairly compact stellar system that managed to run away. As the loss of the extended envelope is related to the tidal stripping and the galaxy never crossed the boundaries of the simulation box, the example is physically grounded and does not represent a numerical artifact \citep{Saulder+2020MNRAS.491.1278S}.

\section{Summary and conclusions} \label{sec:summary}

After the discovery of dark matter-deficient galaxies and the numerous studies that followed, a consensus formed around the notion that these exotic objects could naturally occur in the $\Lambda$CDM model due to stronger tidal interactions through the stripping process. They are typically satellites, with stellar masses in the $10^8-10^9\;\mathrm{M}_\odot$ range, of more massive galaxies. Thus, the recent discovery of a massive galaxy lacking dark matter, without a more massive neighbor, was puzzling. The authors responsible for the discovery discussed two possible scenarios for this to happen, one of which was that the galaxy lost its dark matter early. To test the viability of this scenario, we searched for examples in IllustrisTNG300 cosmological simulation. In this work, we present the evolutionary path of one such example. The first dark matter-poor galaxy in an isolated environment found in cosmological simulations, which is physically grounded and does not represent a numerical artifact.

The galaxy has, at present, a stellar mass of $M_\star \simeq 6.8 \cdot 10^9\; \mathrm{M}_\odot$, no gas, $M_\mathrm{DM}/M_\mathrm{B} \simeq 1.31$, and stellar half-mass radius of $R_{0.5,\star} = 2.45\;\mathrm{kpc}$. While its characteristics do not match those of the NGC 1277 galaxy, its evolutionary path is a perfect example of how a galaxy that loses its extended dark matter and gaseous envelope early, becoming dark matter-poor, can manage to run away from cluster environment and massive neighbors.

The galaxy lost the majority of its dark matter early (starting with a multi-body non-merger encounter), between $z = 2.32$ and $z = 1.53$ (about $9.6$ Gyr ago), even before it formed its present-day stellar component. On its way to the first pericentric passage in the first proto-cluster, around $8.99$ Gyr ago, it accreted a considerable amount of gas, triggering a burst of star formation and sharply increasing the stellar mass to essentially present-day value. While a fraction of accreted gas was then transformed into stars, the majority was stripped between the second and third pericentric passage and the galaxy eventually became gas-free ($\sim 6.67$ Gyr ago). As a consequence, $M_\mathrm{DM}/M_\mathrm{B}$ fluctuates around $1$ until present (i.e., $z=0$).

After the host protocluster merged with a nearby massive galaxy cluster $\sim5.72\;\mathrm{Gyr}$ ago, the galaxy blazed through the new cluster, seemingly unnoticed, with a single pericentric passage. On its way to the apocenter, it had a non-merger interaction with multiple cluster members near the outskirts, then accelerating, diverting its orbit, and eventually leaving the cluster. Since then, for the last $\sim 2\;\mathrm{Gyr}$, it has remained largely unchanged and isolated, residing just outside the most massive cluster in the simulation.

For this scenario to be viable, a galaxy has to lose its extended dark matter envelope very early, that is, before massive galaxy clusters formed. In this way, a galaxy has become compact enough early on (although not necessarily compact by standard definitions and constraints), which has allowed it to avoid mergers in dense environments and to interact with cluster members only through non-merger events. During these interactions, it has been able to accelerate and divert its orbit, eventually ending up leaving the cluster.

The uniqueness of the galaxy presented in this work and its peculiar evolution suggest that the whole evolutionary path, particularly in the later phase, represents a series of particular circumstances. However, circumstantial events, despite being rare, are physically grounded. The upcoming public release of the MillenniumTNG simulations\footnote{\url{https://www.mtng-project.org/}} will primarily expand our knowledge and understanding of large-scale clustering as well as the formation and evolution of galaxy clusters \citep[e.g.,][]{Bose+2023MNRAS.524.2579B,Hernandez-Aguayo+2023MNRAS.524.2556H,Pakmor+2023MNRAS.524.2539P}. As an appealing consequence, it will also provide fertile ground for the study of intercluster dynamics and the evolution of peculiar galaxy classes and events, as well as of environmental effects in general, with much better statistics than have been previously available.

\begin{acknowledgements}
      This research was supported by the Ministry of Science, Technological Development and Innovation of the Republic of Serbia (MSTDIRS) through the contract no. 451-03-47/2023-01/200002 made with Astronomical Observatory (Belgrade, Serbia).
\end{acknowledgements}


\bibliographystyle{aa}
\bibliography{main}

%--------------------------------------------------------------------

\begin{appendix} %First appendix
\section{The first infall:\ A visual example}
\begin{figure}[ht!]
\caption{ Galaxy at $z=2.44$ (snapshot 29) just before the first infall (top) and $z = 2.32$ (snapshot 30) when the infall starts (bottom). Left column shows the stellar density, while the right column shows the gas column density.\\}
\centering \includegraphics[width=0.75\textwidth, keepaspectratio]{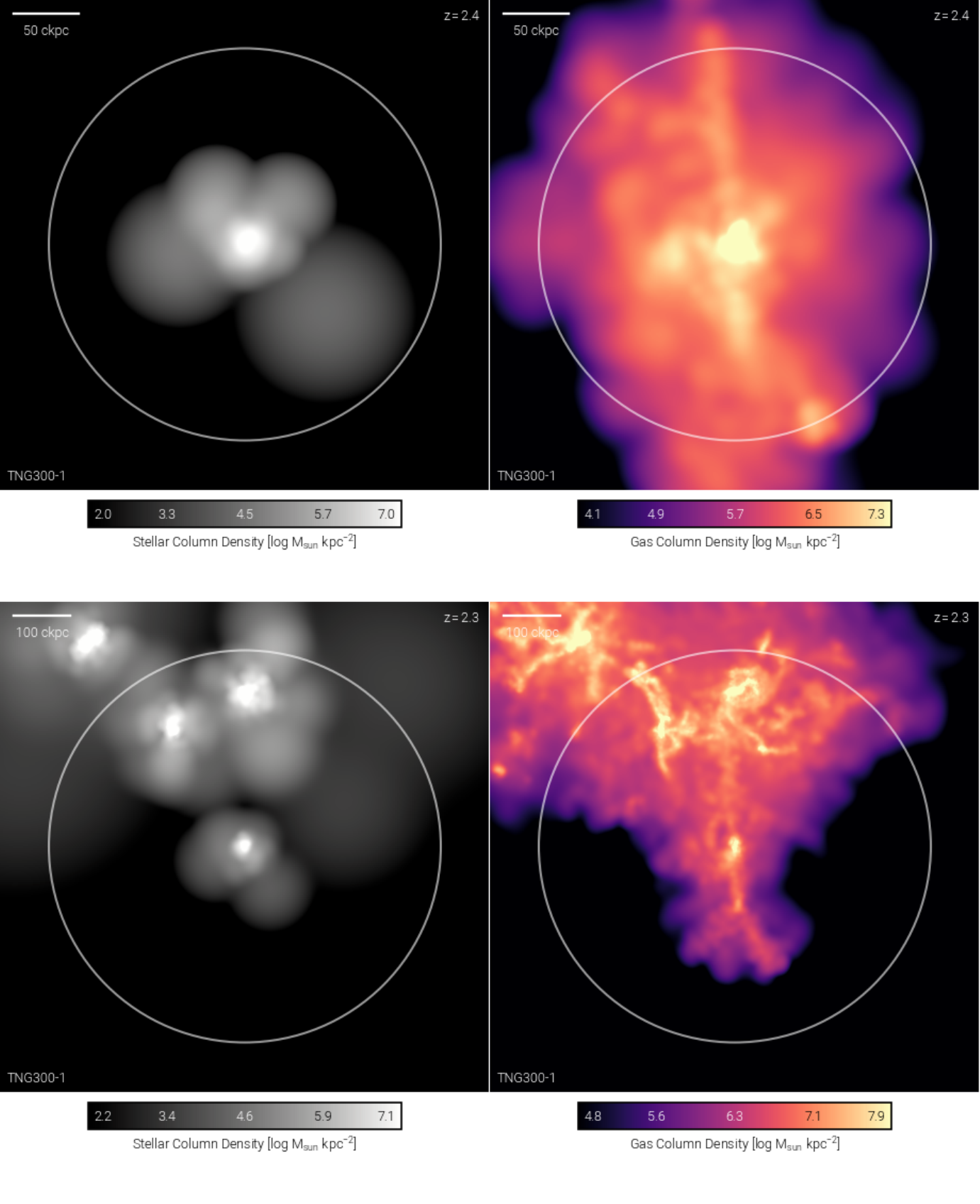}
\label{fig:infall}
\end{figure}

Figure~\ref{fig:infall} was generated using the IllustrisTNG visualization tool\footnote{Available at \url{https://www.tng-project.org/data/vis/}.} and shows the galaxy in two consecutive moments before and after infall begins.

\end{appendix}

\end{document}